\documentstyle[12pt]{article}
\def\lromn#1{\uppercase\expandafter{\romannumeral#1}}

\begin{document}

\begin{titlepage}
\begin{flushright}
August 1995\\
TU/95/488
hep-ph/9508378
\end{flushright}

\vspace{12pt}

\begin{center}
\begin{Large}

\renewcommand{\thefootnote}{\fnsymbol{footnote}}
\bf{
Particle Production and \\
Dissipative Cosmic Field}
\footnote[1]
{Work supported in part by the Grant-in-Aid for Science Research from
the Ministry of \\ \hspace*{0.6cm} Education, Science and Culture of
Japan No. 06640367}

\end{Large}

\vspace{36pt}

\begin{large}
\renewcommand{\thefootnote}{\fnsymbol{footnote}}
H. Fujisaki, K. Kumekawa, M. Yamaguchi, and M. Yoshimura \\
Department of Physics, Tohoku University\\
Sendai 980-77 Japan\\
\end{large}

\vspace{54pt}

{\bf ABSTRACT}
\end{center}

\vspace{0.5cm}
Large amplitude oscillation of cosmic field that may occur
right after inflation
and in the decay process of weakly interacting fields gives rise to
violent particle production via the parametric resonance.
In the large amplitude limit
the problem of back reaction against the field oscillation is solved
and the energy spectrum of created particles
is determined in a semi-classical approximation.
For large enough coupling or large enough amplitude the resulting
energy spectrum is broadly distributed, implying larger production of high
energy particles than what a simple estimate of the reheating
temperature due to the Born formula would suggest.


\end{titlepage}

\vspace{0.5cm}
It is a common belief that coherent field oscillation gives rise to particle
production coupled to the field, and ultimately the field is dissipated
away if no further supply of field energy is provided.
The phenomenon may occur in a number of cosomological situations,
notably in
the reheating stage of inflation and in the decay process of weakly interacting
spinless fields (WISF) such as the Polonyi field in some supergravity models.
Recently it has been recognized
\cite{linde et al 94} \cite{brandenberger et al} \cite{holman 95}
\cite{mine95-1} \cite{reheating parametric}
that presence of the parametric resonance in the
corresponding classical system much enhances the field decay,
in typical cases of
the strong coupling or the large amplitude oscillation the decay proceeding
much more rapidly than what is expected by the Born term.
The effect is important in many ways, and may change our understanding
of the exit problem of inflation and the WISF problem.
For instance, the inflaton decay may rapidly end with non-thermal
particle creation and
the energy redistribution towards the thermal one may take place
by much slower processes of interaction among produced particles.
The new decay mechanism may thus drastically change previous estimate of
the reheating temperature.

In this paper we investigate the back reaction of particle production
against the field oscillation and calculate the energy spectrum of created
particles before effect of particle interaction becomes important.
We do this by fully considering effect of cosmological expansion.
Throughout this work the oscillating field is treated as a classical
variable and the back reaction is incorporated by a semi-classical
approximation.
The formalism of Ref.\cite{mine95-1} based on the
Schr$\stackrel{..}{{\rm o}}$dinger picture of the quantum system
coupled to the field oscillation enables one to solve the problem of
both the back reaction and the particle spectrum in a compact way.
Moreover, we are able to clarify in what sense one may effectively
regard a pure quantum state as a mixed state, even if no interaction
among produced particles is explicitly taken into account.

Inflaton or WISF field is generically denoted in this paper by $\xi (t)$.
It is assumed that after some cosmic time the field undergoes a phase
of the damped oscillation around a stable minimum taken to be
$\xi =0$ here, like
\( \:
\xi (t) \:\propto  \: a^{-3/2}(t)\,\sin \left( m_{\xi }(t-t_{0})\right)
\,,
\: \)
with $m_{\xi }$ the mass of $\xi $ field, and $a(t)$ the cosmic
scale factor.
The $\xi $ field has coupling to various matter and radiation fields, and
we consider a typical Yukawa coupling of massless quantum field
$\varphi $ of the form
\( \:
\frac{1}{2}\, \mu \xi \varphi ^{2} \,.
\: \)
The generic quantum field $\varphi $ may have their own interaction,
and the state of the universe
may be in a thermal equilibrium if the interaction rate is larger than
the Hubble expansion rate.
What we consider in the rest of discussion, however, typically occurs
in time scales shorter than $O(10 - 100)\times 1/m_{\xi }$.
In this circumstance one may envisage that the quantum field $\varphi $
obeys a free field equation with the periodic mass term $\mu \xi (t)$
in the expanding universe.
This should be a reasonable starting point of discussion if one considers
a time duration
much larger than the oscillation time of order $1/m_{\xi }$
and shorter than the interaction time, unless the interaction rate is
much enhanced by enormously large number density of created particles.

Behavior of the $\varphi $ system is unambiguously
described by the density
matrix, which is then decomposed as a direct product of
independent spatial Fourier component $\rho(\vec{k}) $
due to the assumed spatial translational invariance and neglect of interaction.
The bulk of the universe may be in a thermal equilibrium, but
our main interest lies in the state of newly created $\varphi $
particles by the $\xi $ oscillation.
Hence we may concentrate on this part of the density matrix, which
is determined by the wave function of a quantum state.

The wave function of our problem has a Gaussian form of
\( \:
\exp [\,i\frac{\dot{u}}{u}\,q_{\vec{k}}^{2}\,]/\sqrt{u} \,.
\: \)
The function $u(t)$ here obeys the classical equation of harmonic oscillator
with $\omega ^{2} + \mu \xi (t) \,, \omega \equiv |\vec{k}| $
as a time dependent frequency squared.
The initial condition to this solution is specified by a choice of the
corresponding initial quantum state, which we assume
to be the ground state with
the coupling $\mu \xi $ term switched off at $t = t_{0}$, hence
\( \:
u(t_{0}) = (\omega /\pi )^{-1/2} \,, \dot{u}(t_{0})/u(t_{0}) = i\,\omega
\: \).
Take as a convenient basis the Fock space
that diagonalizes $\varphi $ particle number when
the coupling term $\mu \xi $ is switched off.
Non-vanishing density matrix element in each
Fourier component is then given \cite{mine95-1} by
\begin{eqnarray}
&& \hspace*{-1cm}
\rho _{2l \,, 2m} = \frac{1}{\sqrt{\pi }}\,\sqrt{\,\frac{2K}{2+K}\,}\,
\sqrt{\,\frac{\Gamma (l+\frac{1}{2})\,\Gamma (m +\frac{1}{2})}
{l!\,m!}\,}\,(\frac{2-K}{2+K})^{(l+m)/2}\,e^{-2i\delta (l-m)} \,.
\end{eqnarray}
The time dependent real parameters, $K(t)$ and $\delta(t)$,
are related to $u(t)$.
In the small $K$ limit most important in application
($\Re $ and $\Im $ indicating the real and the imaginary parts),
\begin{eqnarray*}
K = \frac{4\omega \Re D}{\omega ^{2} + |D|^{2}} \,, \hspace{0.5cm}
\tan \delta = \frac{\Im D}{\omega } \,, \hspace{0.5cm}
{\rm with} \;
D = (\pi - \frac{i}{2}\,\frac{d}{dt}|u|^{2})/|u|^{2} \,.
\end{eqnarray*}

For a while we ignore effect of cosmological redshift,
focussing on the short time behavior compared to the Hubble time.
With the dimensionless parameters defined by
\( \:
h = \frac{4\omega ^{2}}{m_{\xi }^{2}} \hspace{0.3cm} {\rm and}
\hspace{0.3cm}
\theta = \frac{2\mu \xi }{m_{\xi }^{2}} \,,
\: \)
the classical oscillator equation for $u$ becomes of Mathieu type
\begin{eqnarray}
\frac{d^{2}u}{dz^{2}} + \left( \,h - 2\theta \cos (2z)\,\right)\,u = 0 \,,
\label{mathieu-eq}
\end{eqnarray}
with
\( \:
z = m_{\xi }t/2 \,.
\: \)
The amplitude $\xi $, hence $\theta $ is taken constant in time
for this part of discussion.
As is well known, the classical solution in the parametric resonance
region is asymptotically written as
\( \:
u(t) = e^{\lambda m_{\xi }t/2}\,P(t) \,,
\: \)
with $\lambda $ real and positive, and $P(t)$ a periodic function.
With this form one may use for parameters of the density matrix
\begin{eqnarray}
&&
K = \frac{4\pi \omega }{|u|^{2}}\,\frac{4}{4\omega ^{2} + (\lambda
m_{\xi })^{2}}
\,, \\
&&
\tan \delta = - \frac{\lambda m_{\xi }}{2\omega } -
\frac{1}{2\omega }\,\frac{d}{dt}\ln |P(t)|^{2}
\,. \label{denst-parameters}
\end{eqnarray}
The produced number of particles given by
\( \:
\sum_{l}\,2l\,\rho _{2l \,, 2l}
\: \)
is
\( \:
\sim 1/K \:\propto  \: |u|^{2} \,,
\: \)
and exponentially grows in the parametric resonance region after performing
time average over the quasi-period.
On the other hand, the phase factor $\delta $ that appears in
off-diagonal elements of the density matrix is almost periodic with period of
\( \:
4\pi /m_{\xi }
\: \).
For a sizable $|\theta| \gg  O[1]$ the parametric resonance readily occurs
in a large parameter region of ($\theta \,, h$) space
except in narrow bands of the stability region.
We first wish to improve the analytic formula for large $\theta $ and derive
a form of the density matrix useful for application.

One can envisage the following analogue quantum mechanical model
in order to estimate the growth factor $\lambda $.
The classical equation Eq.\ref{mathieu-eq} is formally
equivalent to the potential
problem in one dimensional quantum mechanics, with the potential given
by
\( \:
\theta \cos (2z)
\: \)
and with the energy $h/2$.
For $h \ll |\theta |$ there is an infinite succession of
large potential barriers.
The difference from the usual eigenvalue problem is that one allows
unbounded solution, since one assumes here a boundary condition at some
$z = z_{0}$ instead of the boundedness condition at $\pm \infty $.
Rather than considering  an infinite series of barriers at once,
one may relate input amplitude to output
amplitude after passing just one barrier and repeat this process $n$ times,
with \( \:
n = [m_{\xi }t/(4\pi )]
\: \) a large integer.

We thus introduce a transfer matrix ${\cal T}$ that connects the input
amplitude to
the left of one barrier (at $-z_{0}$) to the output amplitude
to the right of this barrier (at $z_{0}$)
as follows;
\begin{eqnarray}
{\cal T} \equiv
\Sigma ^{-1}(z_{0})\,\Sigma (-z_{0}) \,, \hspace{0.5cm}
\Sigma (\pm z_{0}) =
\left( \begin{array}{cc}
u_{L}(\pm z_{0}) & u'_{L}(\pm z_{0}) \\
u_{R}(\pm z_{0}) & u'_{R}(\pm z_{0})
\end{array}
\right) \,,
\end{eqnarray}
taking left- and right-movers as the basis of outgoing wave.
$z_{0}$ is assumed positive and large.
$\Sigma (\pm z_{0})$ is a $2 \times 2$ matrix that transforms from
two independent modes
\( \:
\psi _{c}^{\pm }(z) \,, \; \psi _{s}^{\pm }(z) \,,
\: \)
with the properties of
\( \:
\psi _{c}^{\pm }(\pm z_{0}) = 1 \,, \psi_{c} ^{'\pm }(\pm z_{0}) = 0 \,, \;
\psi _{s}^{\pm }(\pm z_{0}) = 0 \,, \psi_{s} ^{'\pm }(\pm z_{0} ) = 1 \,,
\: \)
to the left and right moving modes $u_{L}(z) \,, u_{R}(z)$ at
$z = \pm z_{0}$.
The transfer matrix can be diagonalized by a similarity transformation,
\( \:
{\cal T} = S^{-1}{\cal T}_{D}S \,,
\: \)
giving diagonal elements $t_{\pm }$ with
\( \:
|t_{+}| > 1\hspace{0.3cm} {\rm and} \hspace{0.3cm} |t_{-}| = 1/|t_{+}|<1 \,,
\: \)
both real for an unbounded solution.
One can then clearly identify the growth factor by
\( \:
\lambda = \frac{\ln |t_{+}|}{\pi } \,.
\: \)

Our approximation for large $\theta $ consists of replacing the periodic
barrier by a piece-wise inverted harmonic oscillators: thus for
an interval of
\( \:
|z| < \frac{\pi }{2} \,, \hspace{0.5cm}
\cos (2z) \:\rightarrow  \: 1 - 2z^{2} \,.
\: \)
The approximation gets better as $|\theta |$ increases.
Let
\( \:
\epsilon = \frac{h}{2\theta } - 1 = \frac{\omega ^{2}}{\mu \xi } -1 \,.
\: \)
The left- and right-moving mode functions with the asymptotic behavior of
\( \:
(2\sqrt{\theta }z)^{(-1\mp i\sqrt{\theta }\,\epsilon )/2}\,
\exp (\mp  i\sqrt{\theta }\,z^{2})
\: \)
are written in terms of the standard parabolic cylinder function
$D_{\nu }(z)$ as
\begin{eqnarray*}
u_{L}(z) =
(e^{-i\pi }\theta )^{-(i\sqrt{\theta }\,\epsilon +1)/8}\,
D_{-(i\sqrt{\theta }\,\epsilon + 1)/2}\,\left( \,e^{i\pi /4}\,
2\theta ^{1/4}\,z\,\right)
\,, \hspace{0.5cm}
u_{R}(z) = u^{*}_{L}(z) \,.
\end{eqnarray*}
Despite of the finite range $z_{0} = \pi /2$ the argument of these functions
contain a large factor $\theta ^{1/4}$, hence one may use the asymptotic
form of these functions.
By using analytic properties of the standard parabolic cylinder function
 \cite{whittaker-watson}, one derives
\begin{eqnarray}
&&
\Sigma (-z_{0}) =
\left( \begin{array}{cc}
\beta  & \alpha  \\
\alpha ^{*} & \beta ^{*}
\end{array}
\right)
\left( \begin{array}{cc}
u_{L}(z_{0}) & -u'_{L}(z_{0}) \\
u_{R}(z_{0}) & - u'_{R}(z_{0})
\end{array}
\right)
\,, \\
&&
\alpha  = \frac{\sqrt{2\pi }\,\theta ^{-i\sqrt{\theta }\,\epsilon/4}\,
e^{-\pi \sqrt{\theta }\,\epsilon /4}}
{\Gamma \left( \,(1 + i\sqrt{\theta }\epsilon )/2\,\right)}\,, \hspace{0.5cm}
\beta  = ie^{-\pi \sqrt{\theta }\,\epsilon /2} \,.
\end{eqnarray}

The transfer matrix is further simplified, by introducing a constant $\chi $
by
\( \:
u'_{R}(z_{0}) = \chi\,u_{R}(z_{0})
\: \)
and using $u_{L} = u^{*}_{R}$.
We obtain
\begin{eqnarray}
{\cal T} = \frac{1}{\Im \chi}\,\Im
\left( \begin{array}{cc}
\chi(\alpha e^{2i\delta } + \beta )
& - \chi^{2}\alpha e^{2i\delta } - |\chi|^{2} \beta  \\
-\alpha e^{2i\delta } - \beta
& \chi(\alpha e^{2i\delta } - \beta^{*} )
\end{array}
\right) \,,
\end{eqnarray}
with
\( \:
e^{2i\delta } = u_{R}(z_{0})/u^{*}_{R}(z_{0}) \,.
\: \)
Eigenvalues of the matrix are computed to give
\( \:
|t_{\pm }| = \frac{1}{2}\, |{\rm tr}\,{\cal T}| \pm
\sqrt{\,({\rm tr}\,{\cal T}/2)^{2} - 1\,} \,.
\: \)
The eigenvector corresponding to the growing mode is of great interest.
In the basis of left-right movers the eigenvector
\( \:
\Sigma (z_{0})\vec{v}_{+}
\: \)
has a form of
\( \:
(V\,, V^{*})^{T} \,.
\: \)
The growing mode is thus an equal mixture of the left and the right movers
like
\( \:
-V^{*}u_{L} + Vu_{R} \,,
\: \)
which becomes a single sinusoidal function asymptotically.
Profound implication of this fact will be discussed shortly.

The similarity transformation $S$ that diagonalizes ${\cal T}$
may be written in terms of the eigenvectors
$\vec{v}_{\pm }$.
A large number of repeated barrier crossover leads to
\begin{eqnarray}
&&
{\cal T}^{n} = S^{-1}{\cal T}_{D}^{n}S =
\frac{t_{+}^{n}}{2}\, \left( \begin{array}{cc}
1 & X \\
\frac{1}{X} & 1
\end{array}
\right) +
\frac{t_{-}^{n}}{2}\, \left( \begin{array}{cc}
1 & -X \\
-\,\frac{1}{X} & 1
\end{array}
\right)
\,, \\
&& \hspace*{1cm}
X = -\,{\rm sign}\,(\Re (\alpha e^{2i\delta }))\,
\Im \chi\,\sqrt{\,\frac{\Im (\beta - \alpha e^{2i\delta })}
{\Im (\beta + \alpha e^{2i\delta })}\,}
\,, \label{amplification}
\end{eqnarray}
assuming a purely imaginary $\chi $ relevant later.
The initial state corresponding to the quantum ground state is given by
\( \:
\vec{v}_{0} \:\propto  \: (1 \,, i\tilde{\omega })^{T} \,, \;
\tilde{\omega } = \frac{2\omega }{m_{\xi }} \,,
\: \)
in this basis.
The norm of the state vector is best determined in the basis of the left
and the right movers that have the unit flux.
Comparing the two norms, one finds that
after many barrier crossing the initial state vector $\vec{v}_{0}$
is amplified by
\begin{eqnarray}
t_{+}^{n}\,\frac{1}{2}\, \sqrt{\,\frac{(X^{2}+\tilde{\omega }^{2})
(X^{2}+|\chi| ^{2})}{X^{2}(|\chi| ^{2}+\tilde{\omega }^{2})}\,} \,,
\label{prefactor}
\end{eqnarray}
for $\chi $ purely imaginary.
Besides the usual factor $|t_{+}|^{n} = e^{n\ln |t_{+}|}$, it contains
a prefactor.

One can simplify relevant formulas in the large $\theta $ limit.
First, the constant $\chi $ is of order $\sqrt{\theta }$
and is purely imaginary:
\( \:
\chi = i\sqrt{\theta }\,(2z_{0} + \frac{\epsilon }{2z_{0}}) \,.
\: \)
The factor $X$ in Eq.\ref{amplification} is large and of order
$\sqrt{\theta }$:
\begin{eqnarray*}
X = -\,\sqrt{\theta }\,\left( \pi - \frac{1}{\pi }(1-\frac{h}{2\theta })
\right)\,
\frac{1-\tan (\psi /2)}{1+\tan (\psi /2)} \,.
\end{eqnarray*}
{}From this the extra prefactor in the amplification formula Eq.\ref{prefactor}
turns out of order unity to leading order of $\theta $ and has a complicated
energy dependence.
In the rest of numerical analysis
we neglect this energy dependent prefactor for simplicity.
For $\epsilon < 0$ of our interest the growth factor
\( \:
\lambda \simeq \ln |2\Re (\alpha e^{2i\delta })|/\pi
\: \) is computed as
\begin{eqnarray}
&&
\lambda  = \frac{1}{2}\, \sqrt{\theta }\,(1 - \frac{h}{2\theta })
+ \frac{\ln 2}{\pi } + \frac{1}{\pi}\,\ln |\cos \psi | \,, \\
&&
\psi = \frac{\sqrt{\theta }\epsilon }{2} + \frac{1}{2}\, \pi ^{2}\sqrt{\theta }
- \frac{\sqrt{\theta }\epsilon }{2}\,\ln (\frac{|\epsilon |}{2\pi ^{2}})
\,.
\end{eqnarray}
In the spirit of large $\theta $ approximation we ignore, in what follows,
the constant term $\ln 2/\pi $, and the complicated logarithmic term,
which should be a good approximation except in narrow stability bands of
\( \:
|\cos \psi | < e^{-\pi \sqrt{\theta }|\epsilon| /2} \,.
\: \)
The instability region with $\epsilon > 0$ is very much limited and
we ignore contribution from this region.
Thus the main result in this part of calculation is summarized as
\( \:
\lambda  = \sqrt{\,\frac{\mu \xi}{2}\,}\,
(1-\frac{\omega ^{2}}{\mu \xi })/m_{\xi } \,.
\: \)

The result of the growing mode $\Sigma (z_{0})\vec{v}_{+}$
has an important consequence on behavior of off-diagonal
density matrix elements.
{}From the formula Eq.\ref{denst-parameters}
and the fact that the asymptotic form of the growing mode is
given by a single sinusoidal function,
it is found that the periodic part of the classical solution
$P(t)$ has zeros, hence the factor $\tan \delta $ oscillates between
$\pm \infty $. The oscillation period is small, given by $O[1/m_{\xi }]$.
The time average over a larger time interval then gives vanishing off-diagonal
elements since
\( \:
\rho _{2l \,, 2m} \,\propto  \, e^{-2i\delta (l-m)}
\: \).
Furthermore, neglected interaction among produced particles
is expected to readily break the delicate phase coherence
present in the quantum state of our consideration.
Thus even if the quantum $\varphi $ system is in a pure state in the strict
sense,
it is reasonable to suppose that the system is effectively in a mixed state,
described solely by the diagonal part of the density matrix.
{}From this diagonal density matrix one may compute necessary quantities
such as the average energy density, the energy distribution function,
and more importantly an effective entropy which will be discussed later.
For the rest of discussion we thus use the reduced density matrix
$\rho _{D}$ with off-diagonal elements projected out.

We now consider the cosmological situation in which the amplitude
$\xi $ may vary in time.
In order to discuss the back reaction against the $\xi $ field oscillation,
it is necessary to consider both the cosmological redshift and the dissipation
caused by particle production.
For this purpose we extend the previous analysis, first by
taking an infinitesimally small time period during which cosmological
expansion may be ignored and previous
formulas under periodic perturbation can be applied, and then by incorporating
adiabatic change of the amplitude $\xi (t)$ within  time integral.
Thus we replace the previous factor like
\( \:
\lambda (\xi )\,t
\: \)
by
\( \:
\int_{t_{0}}^{t}\,dt'\,\lambda \left( \xi (t')\right) \,.
\: \)

The $\xi $ field equation with the back reaction included is derived
by considering energy balance between the $\xi $ field and created radiation.
Combined with the Einstein equation, it leads to a  set of equations among the
energy densities, $\rho _{\xi }(t) = \frac{1}{2}\, m_{\xi }^{2}\xi ^{2}(t)$,
 $\rho _{r}(t)$, and the scale factor $a(t)$;
\begin{eqnarray}
&&
\dot{\rho} _{\xi }  + 3\frac{\dot{a}}{a}\,\rho _{\xi } =
 - \frac{d}{dt}\, \langle \,\rho \,\rangle
\,, \\
&&
\dot{\rho}_{r} + 4\frac{\dot{a}}{a}\,\rho _{r} =
\frac{d}{dt}\, \langle \,\rho \,\rangle
\,,  \\
&&
(\frac{\dot{a}}{a})^{2} = \frac{8\pi G}{3}\,(\rho _{\xi } + \rho _{r})
\,,
\end{eqnarray}
where
\( \:
\langle \,\rho \,\rangle = {\rm Tr}\,(\,\omega N_{\omega } \rho _{D}\,) \,,
\: \)
with the trace including the phase space integration.
Because
\( \:
\lambda  = \sqrt{\frac{\mu \xi }{2}}\,(1-\frac{\omega ^{2}}{\mu \xi })/m_{\xi }
\: \),
the source term
\( \:
\frac{d}{dt}\, \langle \,\rho \,\rangle
\: \)
for a single mode has the form of
\begin{eqnarray*}
\sqrt{\mu \xi (t)}\,(1-\frac{\omega ^{2}(t)}{\mu \xi (t)})\,
\exp [\,c\,\int_{t_{0}}^{t}\,dt'\,\sqrt{\mu \xi (t')}\,
(1-\frac{\omega ^{2}(t)}{\mu \xi (t')}\,\frac{a^{2}(t)}{a^{2}(t')})\,] \,,
\end{eqnarray*}
with $\omega ^{2}(t)a^{2}(t)$ independent of time.
Somewhat by accident, the phase space integral involved here can be
performed explicitly in the form of
\( \:
\int_{0}^{\omega _{c}}\,d\omega \,\omega ^{2n+1}\,e^{-C\omega ^{2}}
\: \)
with $n=1\,, 2$.
This yields a closed set of differential equations
instead of complicated integro-differential equations.
In terms of dimensionless quantities divided by the initial $\xi $ energy
density $\rho _{\xi }^{0} = \frac{1}{2}\, m_{\xi }^{2}\xi _{0}^{2}\,, $
\( \:
Y \equiv  \rho _{\xi }/\rho _{\xi }^{0} \,, \;
Z \equiv \rho _{r}/\rho _{\xi }^{0} \,, \;
\tau \equiv m_{\xi }t \,,
\: \)
these are \\ ($'$ indicating $\frac{d}{d\tau }$)
\begin{eqnarray}
&&
\hspace*{-1cm}
U' = c\sqrt{g_{0}}\,Y^{1/4}\,U \,,\hspace{0.5cm}
W' = c\sqrt{g_{0}}\,Y^{1/4} + \frac{Y'W}{2Y} +2\frac{a'}{a}\,W
\,, \hspace{0.5cm}
\frac{a'}{a} = d\sqrt{Y+Z} \,,
\\
&&
Y' + 3d\,\sqrt{Y+Z}\,Y = - \,B\,Y^{5/4}\,U\,
[\,W - 2 + e^{-W}\,(W + 2)\,]/W^{3} \,, \\
&&
Z' + 4d\,\sqrt{Y+Z}\,Z = B\,Y^{5/4}\,U\,
[\,W - 2 + e^{-W}\,(W + 2)\,]/W^{3}  \,,
\end{eqnarray}
where
\( \:
g_{0} = \frac{\mu \xi _{0}}{m_{\xi }^{2}} = \theta _{0}/2 \,, \;
d = \sqrt{\frac{4\pi }{3}}\,\frac{\xi _{0}}{m_{{\rm pl}}} \,, \;
B = \frac{c}{2\pi ^{2}}\,g_{0}^{5/2}\,(\frac{m_{\xi }}{\xi _{0}})^{2}
\,,
\: \)
with $c $ a numerical factor of order unity:
\( \:
c \sim  1/\sqrt{2} \,.
\: \)
By rescaling the time variable, one sees that this system of equations
has two intrinsic parameters,
\( \:
c^{2}g_{0}/d^{2}  \; {\rm and} \; B/d \,,
\: \)
which we call initial amplitude and damping factor in what follows.

We first discuss asymptotic behavior of solution
that can be worked out by analytic means.
A consistency check shows that the $\xi $ and the radiation energy density
asymptotically behaves like
\begin{eqnarray*}
\rho _{\xi }  \:\rightarrow  \: \rho _{\xi }^{0}\,
(c\sqrt{g_{0}}\,\tau \ln (\tau
/\tau _{c}) )^{-4}
\,, \hspace{0.5cm}
\rho _{r} \:\rightarrow  \: \rho _{\xi }^{0}/(2d\,\tau )^{2}
\,, \hspace{0.5cm}
a'/a \:\rightarrow \: 1/(2\tau) \,,
\end{eqnarray*}
where $\tau _{c}$ is a constant determined from early time behavior.
The created radiation density $\rho _{r}$
asymptotically follows the standard time variation
of radiation dominated universe, but
variation of the $\xi $ energy density
is balanced against both the redshift and the particle creation terms.
Unless some other mechanism works, the $\xi $ field settles down to
this self-decaying configuration, whose energy density
is however much more suppressed than that of the non-relativistic
matter $\:\propto  \: a^{-3}$.
As will be discussed shortly, this does not mean that the particle
creation permanently lasts, however.

It is now appropriate to discuss how the resonant particle production
ends and what the fate of the $\xi $ field oscillation would be.
As the $\xi $ amplitude decreases both by cosmological damping and particle
production, the $\theta $ parameter given by
$2\mu \xi /m_{\xi }^{2}$ also decreases, and when $\theta $ becomes
comparable to unity, relevant mode may enter into the small amplitude regime,
or even into a stability band region.
We suppose that
the strong resonant decay effectively ends at this time
and the $\xi $ matter behaving like a non-relativistic matter persists
until the Born decay lifetime becomes comparable to the Hubble time.
Note in this context that
within the first resonance band region the small amplitude decay rate formula,
when translated per one $\xi $ particle by dividing the number density
\( \:
n_{\xi } = \frac{1}{2}\, m_{\xi }\xi ^{2} \,,
\: \)
agrees  \cite{mine95-1}
with the Born formula $\mu ^{2}/(32\pi m_{\xi })$.
This means that the small amplitude region is well approximated by the
Born formula, although the intermediate amplitude region of
\( \:
\theta \sim 1
\: \)
is rather difficult to deal with.

Let then the end time of resonant decay be denoted by
\( \:
\tau = \tau _{e} \,.
\: \)
After this time and before the Born decay starts, the $\xi $ matter density
behaves like
\( \:
\rho _{\xi }/\rho _{r} \:\propto  \: a(\tau )/a(\tau _{e}) \,.
\: \)
The Born decay starts around
\( \:
\tau = \tau _{d} = 16\pi m_{\xi }^{2}/\mu ^{2} \,,
\: \)
from which one may estimate the fraction of $\xi $ matter density at
the epoch of Born decay,
\begin{eqnarray}
\frac{\rho _{\xi }}{\rho _{r}} =
16\sqrt{\pi }\,f^{3/2}\,d^{1/2}\,\frac{m_{\xi }}{\mu \theta  _{0}^{2}}
= 4\sqrt{2\pi }\,(\frac{\pi }{3})^{1/4}\,f^{3/2}\,
\frac{m_{\xi }^{5}}{\mu ^{3}\xi _{0}^{3/2}m_{{\rm pl}}^{1/2}}
\,,
\end{eqnarray}
with
\( \:
f = d\tau _{e}
\: \)
of order unity numerically.
This gives an upper bound on the parameter combination,
\( \:
m_{\xi }^{5}/(\mu ^{3}\xi _{0}^{3/2}m_{{\rm pl} }^{1/2}) =
\xi _{0}^{3/2}/(g_{0}^{3}m_{\xi }m_{{\rm pl}}^{1/2})
\,,
\: \)
in order not to have a $\xi $ dominant epoch.
In the rest of discussion we assume that this bound is obeyed.
Namely, we are considering a situation in which
the $\xi $ dissipation initially
takes place via the rapid resonant particle production and at a later epoch
a small $\xi $ residual completely disappears via the Born decay.
Details of this mechanism, in particular how individual Fourier modes undergo
the Born phase of decay are left to further work.
In actual situations there may well exist further strong
dissipation of the $\xi $ field due to the
intermediate amplitude oscillation around $\theta =1$.
Here we simply assume that there is no sizable late time decay and
ignore a possible spectrum component due to the Born decay.

Details of the particle energy spectrum are obtained by solving the Boltzmann
equation.
A Boltzmann-like equation in our model of the expanding universe
may be set up for the distribution
function $n(\omega \,, t)$ of produced particles, once the source term is
given by the reduced density matrix;
\begin{eqnarray}
\omega \frac{\;\partial}{\partial t} \,n(\omega \,, t) -
\frac{\dot{a}}{a}\,\omega ^{2}\,\frac{\partial }{\partial \omega }\,
n(\omega \,, t) = \omega S[\xi ] \,,
\end{eqnarray}
with
\( \:
S[\xi ]  = \frac{d}{dt}\,{\rm tr}\;(\,N_{\omega }\rho _{D}\,) =
\frac{d}{dt}\, \sum_{l}\,2l\,\rho _{2l\,, 2l}\,.
\: \)
The distribution function $n(\omega \,, t)$ here,
when integrated over the entire phase space,
is not normalized to unity as usual. It rather gives the number density
after the phase space integration.
Solution to the Boltzmann-like equation is then
\begin{eqnarray}
n(\omega \,, t) = \int_{t_{0}}^{t}\,dt'\,S\left( \omega \frac{a(t)}{a(t')}
\,, t' \right) \,,
\end{eqnarray}
from which one can compute the spectrum.

Result of numerical computation is presented in a few figures.
For lack of space we only show typical parameter cases of
\( \:
B/d = 10^2 \,, 1 \,, 10^{-2} \,,
\: \)
and
\( \:
c\sqrt{g_{0}}/d = 10 \,,
\: \)
which we call large, intermediate, and small damping cases, all taking
a large initial amplitude, $\theta _{0} = 200$.
As the initial condition to the differential equation we take for
simplicity the $\xi $ dominance so that the initial radiation energy
density
\( \:
\rho _{r}(t_{0}) = 0
\: \)
with the Hubble rate
\( \:
H(t_{0}) = m_{\xi }\,.
\: \)
The time scale in these figures is
\( \:
1/(d\,m_{\xi }) = \sqrt{\frac{3}{4\pi }}\,\frac{m_{{\rm pl}}}
{\xi _{0}m_{\xi }} \,.
\: \)
We take this time scale larger than $1/m_{\xi }$ so that the dissipation
time scale is much larger than the field oscillation period, for consistency
of our approximation.
In Fig.1 is shown time variation of the $\xi $ and the net radiation
energy density,
\( \:
\rho _{\xi }(t) \,, \; \rho _{r}\frac{a^{4}(t)}{a^{4}(t_{0})} \,,
\: \) along with the scale factor $\frac{a(t)}{a(t_{0})}$
for the intermediate damping.
Note that
the asymptotic behavior is observed already at moderately early times
of a few times $1/(dm_{\xi })$.
In Fig.2 is shown time variation of the energy spectrum in unit of
$m_{\xi }^{3}/(2\pi ^{2})$ plotted against the energy in unit of
$\sqrt{\mu \xi _{0}}$, in which is marked by the solid line
the spectrum at the exit time of resonant particle production,
taken for a guide to occur when $\theta = 1$, giving $d\tau \sim 1.2$ for
this parameter set. The peak position of the spectrum shifts towards the
low energy side as time proceeds, but the maximal peaking occurs at a time
slightly earlier than the exit time.
In Fig.3 we show the energy spectrum at the exit time
in the three different cases of damping.
The total number of created particles increases as the damping factor
$B/d$ decreases.

Although the number distribution at the exit time is not the thermal one,
there is a broad range of energies of which particles are produced,
especially for larger initial amplitudes and larger damping.
This tends to accelerate the approach to the thermal distribution
in subsequent energy exchange process by interaction among created particles.
The broadness of energy distribution
is more pronounced for larger damping or larger initial
amplitudes, as illustrated in Fig.3.
What is important to physical processes that occur at later times is
the average energy and energy dispersion.
We found numerically
that in the dimensionless unit the energy dispersion approaches
1,
\( \:
\sqrt{\,\langle \omega ^{2} \rangle - \langle \omega  \rangle^{2}\,}/
\langle \omega  \rangle \:\rightarrow  \: 1
\: \)
in the small $B$ limit. This is a good measure of the broadness of
the energy distribution.

A quantitative measure of dissipation is given by the entropy.
Definition of the entropy in a classical system is somewhat ambiguous.
In our approach based on the quantum behavior of radiation system it is natural
to consider the quantum statistical entropy
determined by the reduced density matrix $\rho _{D}$.
One may thus define the entropy of created particles by
\( \:
s(t)\,V = -\,{\rm Tr}\,(\rho _{D}\,\ln \rho _{D}) \,,
\: \)
again the big trace including the phase space integral.
In general it is not easy to derive a tractable expression of the
entropy from this general formula,
but use of the asymptotic form of the diagonal density matrix
element for large $l$,
\( \:
\rho _{2l \,, 2l} \sim \sqrt{\frac{K}{\pi }}\,\frac{1}{\sqrt{l}}\,e^{-Kl}\,,
\: \)
yields a simple, but suggestive form of
\( \:
s_{\omega } \sim \ln \langle N_{\omega } \rangle
\: \)
for a single mode, ignoring a constant term of order $0.1$.
In terms of the previous variables, mode-summed entropy density at time $t$
is then
\begin{eqnarray}
s(t) =
\frac{g_{0}^{2}}{2\pi ^{2}}\,m_{\xi }^{3}\,\frac{2}{15}\,a^{-3}(t)\,
\int_{\tau _{0}}^{\tau }\,d\tau '\,Y(\tau ')\,a^{3}(\tau ')
\,.
\end{eqnarray}
The entropy computed according to this formula is plotted in Fig.4.
It is seen that after initial transient behavior ends,
the entropy decreases like $a^{-3}(t)$ with the scale factor, as expected.
The asymptotic values of the net entropy $s(t)a^{3}(t)/a^{3}(t_{0})$
are
\( \:
4.0 \,, 43 \,, 110
\: \)
in unit of $m_{\xi }^{3}$ for the large, intermediate, and small damping.

For cosmological application it is important to be able to determine
the reheat temperature after inflation.
Even for a broad energy distribution of produced particles,
the energy redistribution towards the thermal one may be established only
after a long elapse of time due to a small coupling.
Nevertheless it might be reasonable, as a crude approximation in the case of
broad particle distribution,
to estimate the reheat temperature $T_{R}$ by
equating the radiation energy density at the exit time to
\( \:
\frac{\pi ^{2}}{30}\,T_{R}^{4} \,,
\: \)
the usual thermal energy density at the temperature $T_{R}$.
This estimate gives
\begin{eqnarray}
T_{R} = (\frac{15}{\pi ^{2}})^{1/4}\,Z^{1/4}(\tau _{e})\,
\sqrt{\xi _{0}m_{\xi }}
= \frac{1}{\pi }\,(\frac{15}{2\sqrt{2}})^{1/4}\,g_{0}^{5/8}\,
B^{-1/4}\,Z^{1/4}(\tau _{e})\,m_{\xi } \,,
\end{eqnarray}
which in cases of most interest yields
a fraction times $\sqrt{\xi _{0}m_{\xi }}$.
A similar estimate is obtained by assuming instantaneous thermalization
and equating the Hubble rate
\( \:
H = \sqrt{\frac{8\pi ^{3}N}{90}}\,\frac{T_{R}^{2}}{m_{{\rm pl}}}
\: \)
to the inverse of the exit time $1/(2t_{e})$, again giving
\( \:
T_{R} =
\: \) a fraction times $\sqrt{\xi _{0}m_{\xi }}$.
The reheat temperature thus derived is larger by a big margin than
that given by the Born decay rate.
The enhancement factor is roughly of order
\( \:
O[10]\,\frac{m_{\xi }}{\mu }\,\sqrt{\frac{\xi _{0}}{m_{{\rm pl}}}} \,.
\: \)
For instance, in a chaotic type of inflation with
\( \:
\mu = 10^{-4}\,m_{\xi } \,, \hspace{0.3cm} m_{\xi } = 10^{-6}\,m_{{\rm pl}}
\,, \hspace{0.3cm} \xi _{0} = O[m_{{\rm pl}}] \,,
\: \)
this factor may become as large as $10^{5}$, and
\( \:
T_{R} \sim \sqrt{\xi _{0}m_{\xi }} = O[\,10^{16}\,]\,{\rm GeV}
\: \).
But in this particular example the energy spectrum is peaked around
a low value of
\( \:
\omega /\sqrt{\mu \xi _{0}} \,.
\: \)
Indeed, the concept of the reheating temperature is dubious for cases
in which
\( \:
T_{R}
\: \)
is much larger than the kinematical limit of $\sqrt{\mu \xi _{0}} =
\sqrt{g_{0}}\,m_{\xi }$.
The example of the chaotic inflation exceeds this limit.
What happens for very small $B$
is that the total energy density is not a good measure of
the reheat temperature, because an enormous amount of particles is
produced to give a large energy density. The distribution itself
is broad, as evidenced by the unit energy dispersion and not too small
an average energy.

With a larger reheat temperature or more precisely with copious production
of high energy particles of GUT scale energies,
baryo-genesis may proceed efficiently
in the original GUT scenario of asymmetry generation.
A new serious problem would be a copious production of gravitino in
supergravity models. The usual constraint \cite{gravitino problem}
on the reheat temperature favored by low energy supersymmetry
is not met here due to the large number density of created particles.
Resolution of this problem is left to future.

How about the Polonyi problem \cite{polonyi}?
We must admit that our present work does not deal with the intermediate
case of $\theta \sim 1$. Without considering a possible strong dissipation
in the intermediate amplitude region the Polonyi field is damped by the factor
\( \:
\frac{3}{16\pi }\,(\frac{m_{{\rm pl}}}{\xi _{0}})^{2}\,\frac{1}{g_{0}^{2}}
\: \)
due to the resonant particle production. This is presumably not sufficient
to solve the Polonyi problem, but we caution that the intermediate
amplitude region must be included for a definite conclusion.

In summary, we showed in this paper that cosmic field oscillation such as
inflaton or the Polonyi field is
dissipated via rapid resonant particle production if the initial amplitude
or the coupling to matter is large enough.
The energy spectrum of produced particles along with detailed evolution
of the field energy density may be followed by solving a closed set of
differential equations in the large amplitude limit.
Much larger production of high energy particles becomes possible than
what the usual Born decay would imply. This may have great
impact on baryo-genesis and gravitino production.

\newpage

\begin{Large}
\begin{center}
{\bf Figure caption}
\end{center}
\end{Large}

\vspace{1cm}
\hspace*{-0.5cm}
{\bf Fig.1}

Time evolution of $\xi $ energy density (dash-dotted line)
and radiation energy density times scale factor to the 4th power
(solid line) together with
the Hubble rate $\dot{a}/a$ (broken)
for the intermediate set of parameters of
\( \:
c\sqrt{g_{0}}/d =10 \,, B/d =1 \,.
\: \)
The time scale in this and Fig.4 is
\( \:
\sqrt{\frac{3}{4\pi }}\,\frac{m_{{\rm pl}}}{\xi _{0}m_{\xi }} \,.
\: \)
The vertical scale is drawn in arbitrary units, with the $\xi $ energy density
given by the logarithmic scale on the right.

\vspace{0.5cm}
\hspace*{-0.5cm}
{\bf Fig.2}

Time evolution of the energy spectrum in unit of $m_{\xi }^{3}/(2\pi ^{2})$
plotted against the energy in unit of
\( \:
\sqrt{\mu \xi _{0}}
\: \)
for the intermediate damping case, as in Fig.1.
The solid line is the spectrum at the exit time around
\( \:
\tau d = 1.2 \,,
\: \)
and the peak position shifts towards the low energy side as time
proceeds according to
\( \:
\tau d = 0.5 \,, 0.75 \,, 1 \,, 1.5 \,, 2 \,,
\: \)
as indicated by dash-dotted and dotted lines.

\vspace{0.5cm}
\hspace*{-0.5cm}
{\bf Fig.3}

Energy spectrum in unit of $m_{\xi }^{3}/(2\pi ^{2})$
at various damping factors of
\( \,
B/d = 10^{2} ({\rm dash-dotted}) \,,
\: \)
\(
B/d = 1 ({\rm solid})  \:
\: \) reduced by $10^{2}$, and
\( \,
B/d = 10^{-2} ({\rm broken})
\, \)
reduced by $10^{4}$,
plotted against the energy in unit of
\( \:
\sqrt{\mu \xi _{0}} \,.
\: \)

\vspace{0.5cm}
\hspace*{-0.5cm}
{\bf Fig.4}

Time evolution of the net entropy,
entropy density times $a^{3}(t)/a^{3}(t_{0})$,
for the same damping factor as in Fig.3.


\newpage

\begin{thebibliography}{99}

\bibitem{linde et al 94}
L. Kofman, A. Linde, and A.A. Starobinsky, {\sl Phys.\ Rev.\ Lett.\ }{\bf 73},
3195(1994).

\bibitem{brandenberger et al}
Y. Shtanov, J. Traschen, and R. Brandenberger, {\sl Phys.\ Rev. }{\bf D51},
5438(1995).

\bibitem{holman 95}
D. Boyanovsky, M. D'Attanasio, H.J. de Vega, R. Holman, D.-S. Lee, and \\
A. Singh, preprint PITT-09-95;
{\sl Phys.\ Rev. }{\bf D51}, 4419(1995) and references therein.

\bibitem{mine95-1}
M. Yoshimura, {\em
Catastrophic Particle Production under Periodic Perturbation},
Tohoku preprint, TU/95/484 and hep-th/9506176.

\bibitem{reheating parametric}
A.D. Dolgov and D.P. Kirilova, {\sl Sov.\ J.\ Nucl.\ Phys.\ }{\bf 51},
172(1990);\\
J.H. Traschen and R.H. Brandenberger, {\sl Phys.\ Rev. }{\bf D42},
2491(1990).

\bibitem{whittaker-watson}
E.T. Whittaker and G.N. Watson, {\em A Course of Modern Analysis},\\
(Cambridge University Press, London, 1962),
p348.

\bibitem{gravitino problem}
J. Ellis, J.E. Kim, and D.V. Nanopoulos, {\sl Phys.\ Lett.\ }{\bf 145B},
181(1984); \\
M. Kawasaki and T. Moroi, {\sl Prog.\ Theor.\ Phys.\ }{\bf 93}, 879(1995)
and references therein.

\bibitem{polonyi}
G.D. Coughlan, W. Fischler, E.W. Kolb, S. Raby, and G.G. Ross,
{\sl Phys.\ Lett.\ }{\bf 131B}, 59(1983).


\end{thebibliography}
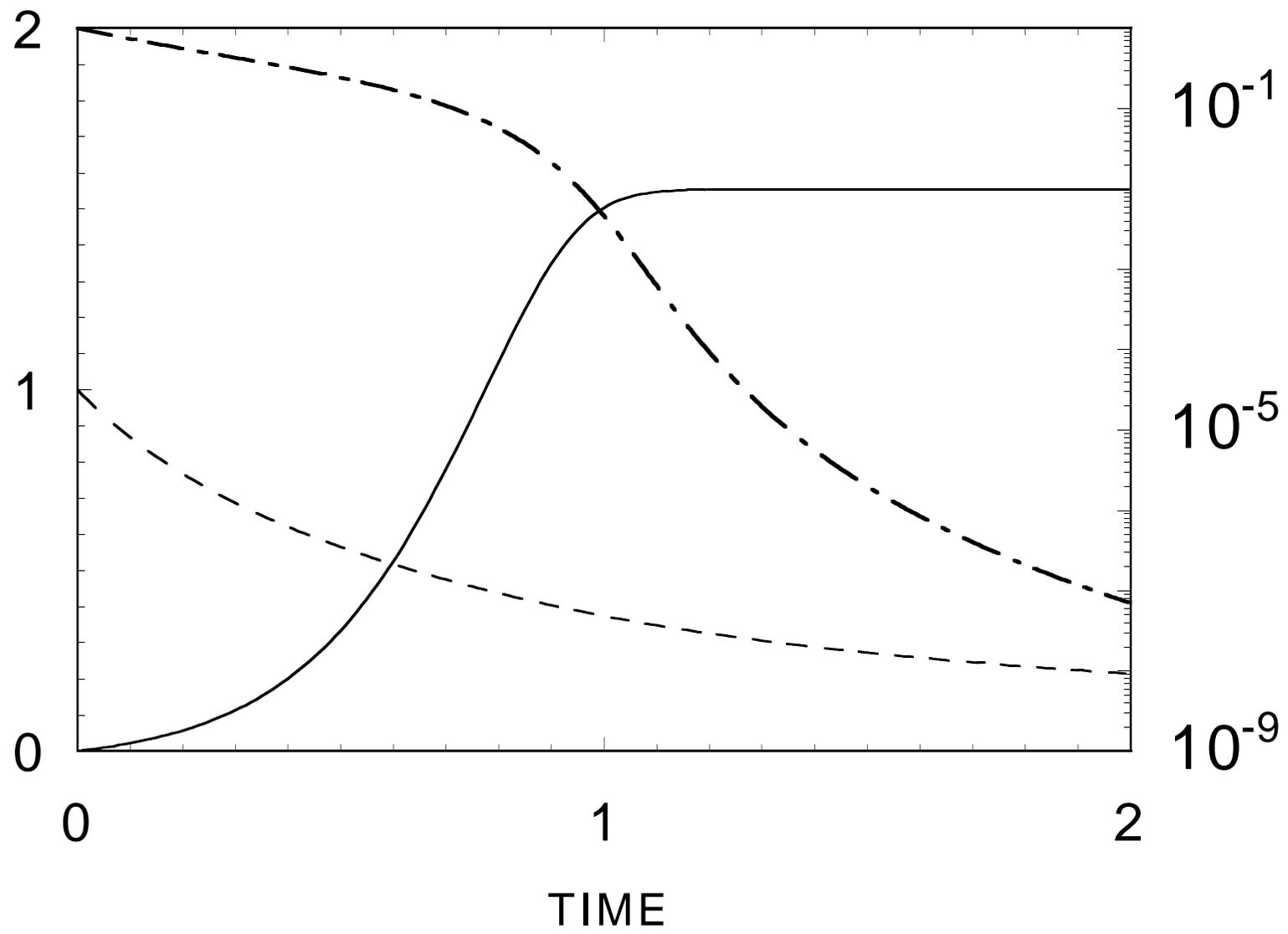
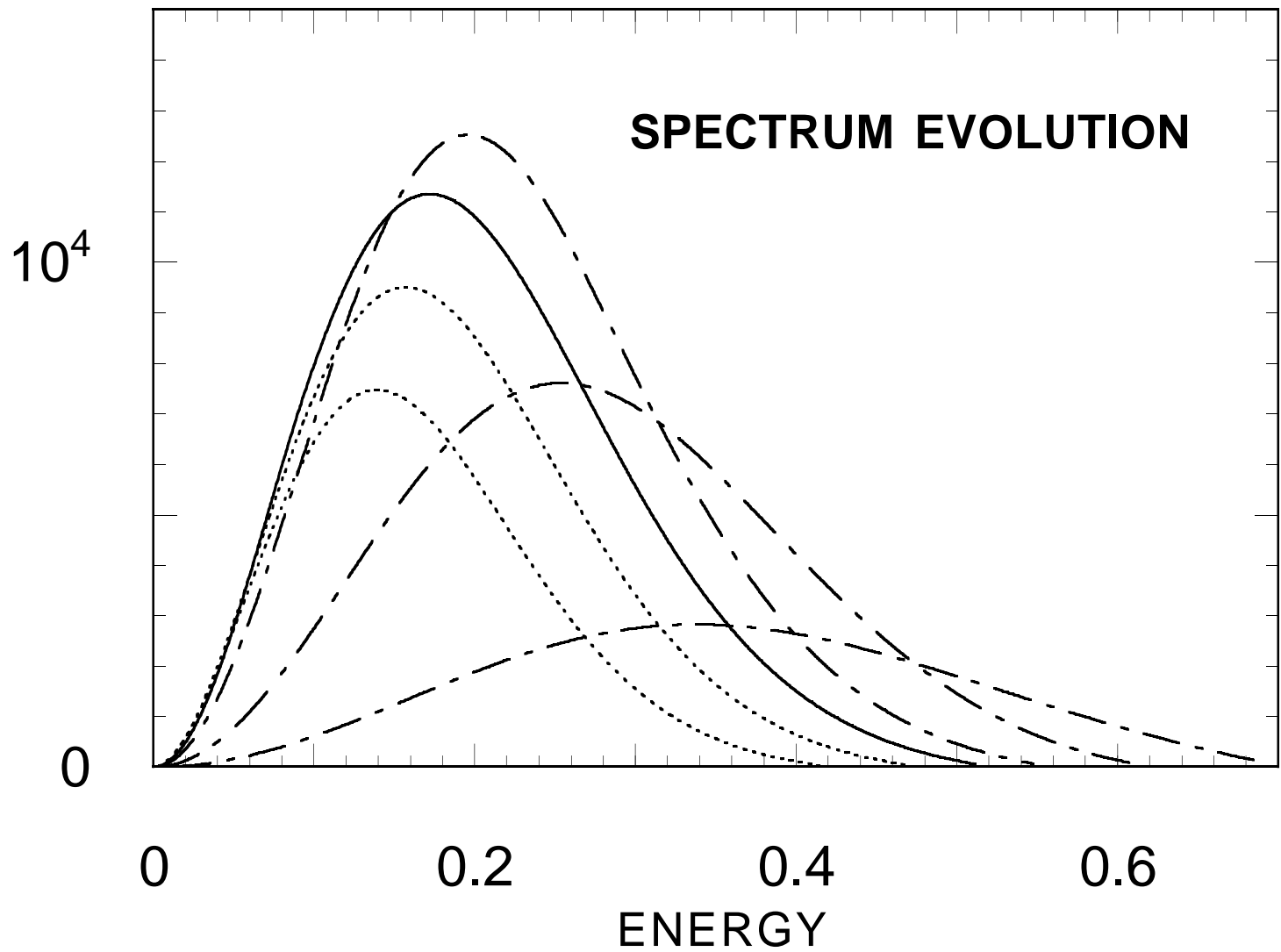
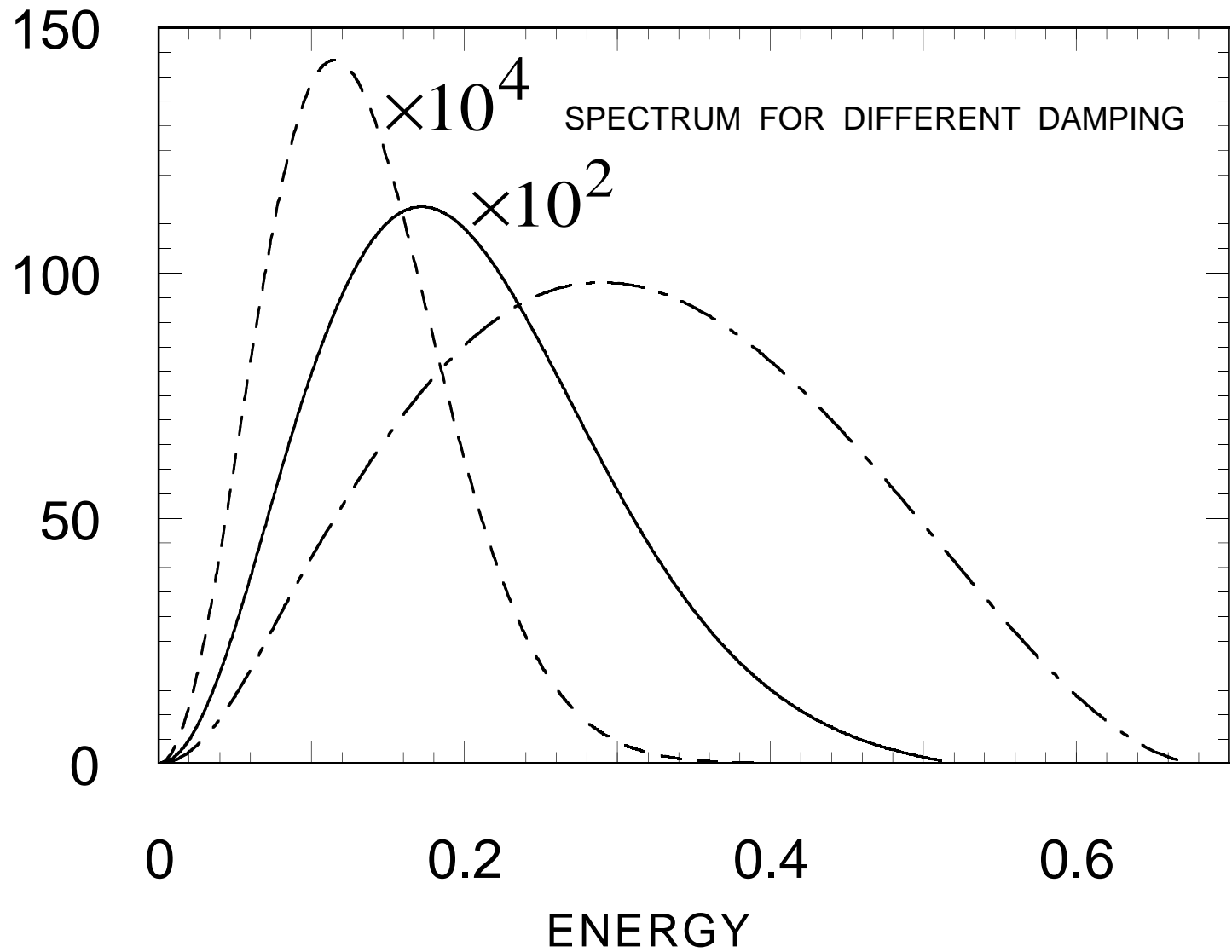
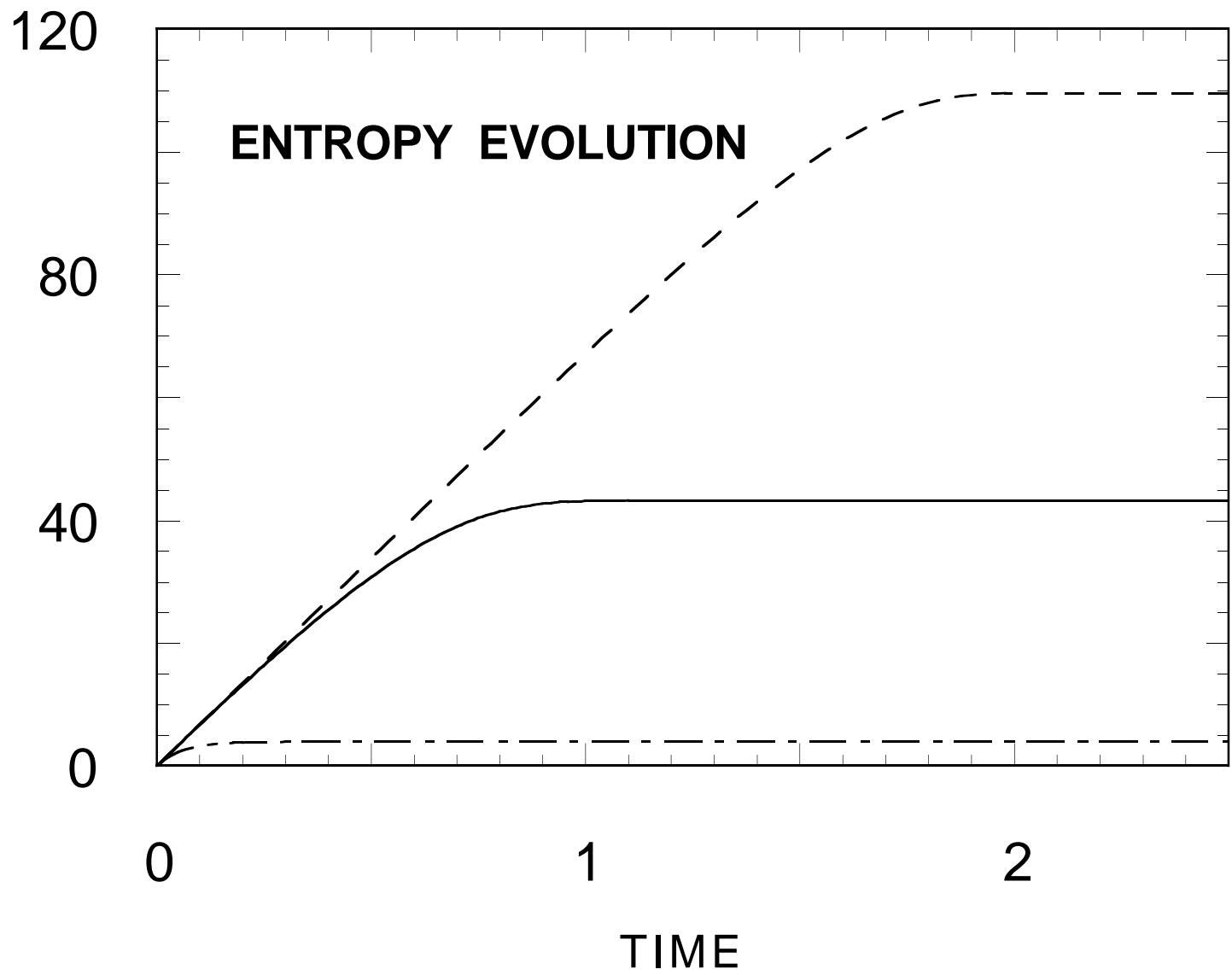
\end{document}